\documentclass{ws-procs961x669}            
\usepackage{ws-toc,ws-multind}
\usepackage{amsmath,amssymb}
\usepackage{color}
\usepackage{aas_macros}

\newcommand{\be}{\begin{equation}}
\newcommand{\en}{\end{equation}}

\def\ltsima{$\; \buildrel < \over \sim \;$}
\def\lsim{\lower.5ex\hbox{\ltsima}}
\def\gtsima{$\; \buildrel $\geq$ \over \sim \;$}
\def\gsim{\lower.5ex\hbox{\gtsima}}

\def\deg{\mbox{$^{\circ}$}}

\begin{document}


\title{Ultra-long period compact sources \\ \small A glimpse into observational breakthroughs and theoretical challenges}

\author{Francesco Coti Zelati$^{1,2}$ and Alice Borghese$^{3}$}

\address{$^1$Institute of Space Sciences (ICE, CSIC),\\ 
Campus UAB, Carrer de Can Magrans s/n, E-08193, Barcelona, Spain\\
E-mail: cotizelati@ice.csic.es}

\address{$^2$Institut d'Estudis Espacials de Catalunya (IEEC),\\ 
08860 Castelldefels (Barcelona), Spain}

\address{$^3$INAF -- Osservatorio Astronomico di Roma,\\
via Frascati 33, I-00078 Monte Porzio Catone (RM), Italy\\
E-mail: alice.borghese@gmail.com}

\begin{abstract}
At the Seventeenth Marcel Grossman meeting, researchers gathered to discuss significant advances in the study of ultra-long period sources. Presentations covered key aspects, including emission properties, evolutionary scenarios, and models for their emission. 
In this proceeding, we summarize key observational breakthroughs and touch upon the proposed evolutionary pathways and state-of-the-art models that seek to explain these sources. Finally, we outline future directions, including the potential of ongoing and upcoming surveys, improved detection algorithms, and multiwavelength observations to significantly expand the known population of these mysterious sources.
\end{abstract}

\keywords{Ultra-long period; Galactic radio sources; radio transient sources; radio bursts; radio pulsars; neutron stars; magnetars; white dwarf stars; fallback discs.}

\bodymatter

\vspace{-0.05cm}
\section{Introduction}\label{sec:intro}
Pulsars are rotating neutron stars (NSs) emitting beams of electromagnetic radiation, typically in the radio band, from their magnetic poles. As these beams sweep our line of sight, they produce regular pulses, making pulsars invaluable for probing phenomena ranging from the interstellar medium and the fundamental properties of matter under extreme conditions to testing predictions of general relativity and theories of alternative gravity. Pulsars are characterized by short spin periods (milliseconds to seconds) and strong magnetic fields in the range of 10$^8$--10$^{15}$\,G\footnote{\url{https://www.atnf.csiro.au/research/pulsar/psrcat/}} \cite{Manchester2005}.

Pulsar radio emission stems from complex magnetospheric processes where charged particles accelerate along magnetic field lines (see \cite{harding2018} for a review). A key feature of this emission is its coherence, meaning that the electromagnetic waves produced by the accelerated particles radiate in phase, substantially amplifying the radio signal. This coherence is responsible for several distinctive observational characteristics of pulsars, including high brightness temperatures, rapid variability, narrow spectral features, and strong polarization.

\begin{figure*}[!h]
\begin{center}
\includegraphics[width=0.9\textwidth]{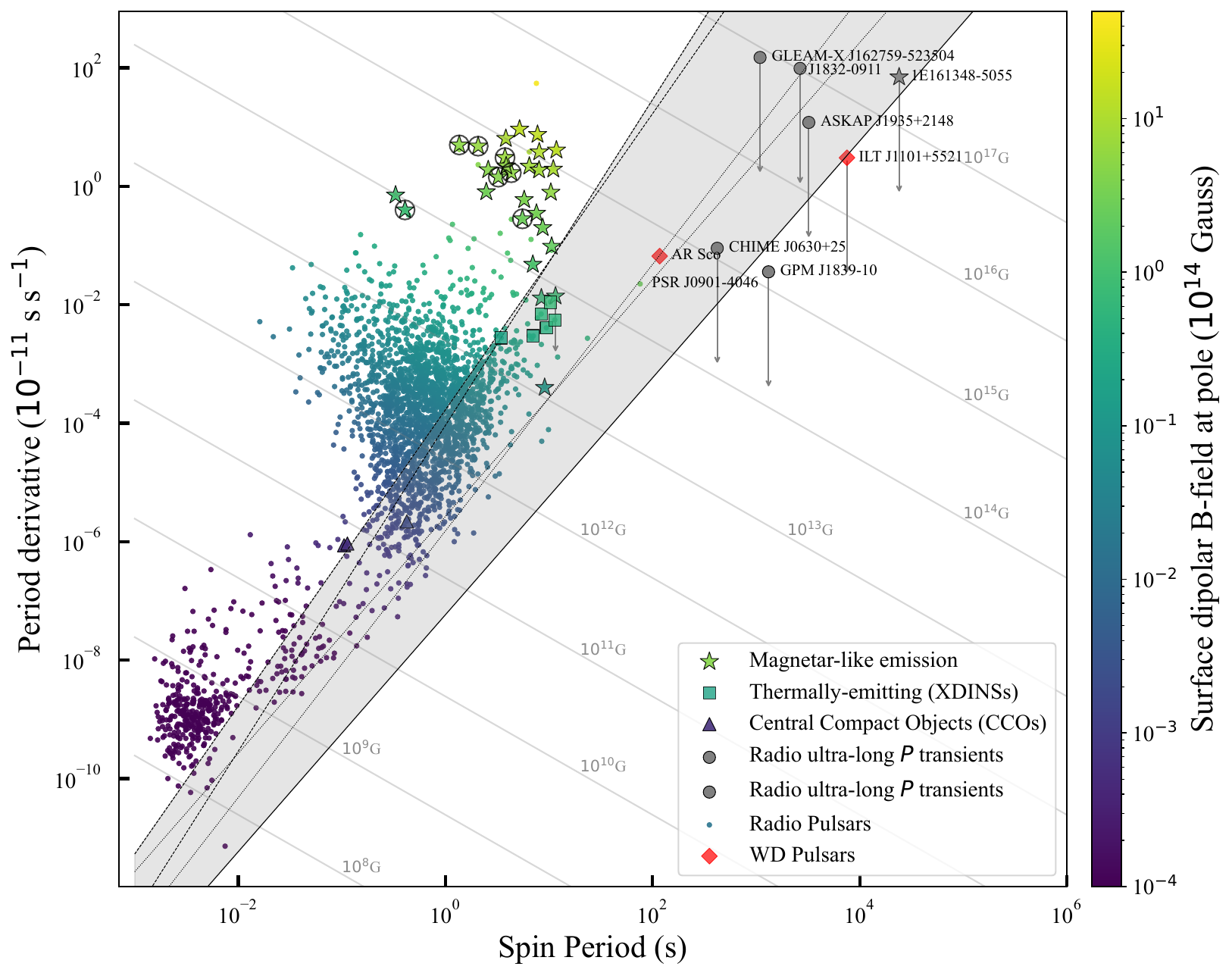}
\caption{\footnotesize{$P-\dot{P}$ diagram for NS and WD pulsars. The red diamonds denote WD pulsars, the gray star marks 1E\,161348$–$5055, the gray circles indicate ULP sources. Arrows represent upper limits on the period derivative. The solid gray lines correspond to lines of constant magnetic field, estimated at the pole. The grey-shaded region denotes the death valley for coherent pulsed radio emission \cite{Chen1993}. The dashed
lines correspond to the theoretical death lines for a pure dipole, dotted lines for a twisted dipole and solid lines for the twisted multipole configuration. The plot was created adapting the code from https://github.com/nhurleywalker/GPMTransient.}} 
\label{fig:ppdot}
\end{center}
\end{figure*}

Pulsars are primarily discovered through large-scale radio sky surveys. Notable examples include the FAST Galactic Plane Pulsar Snapshot \citep{Han2021}, \emph{Arecibo} Pulsar ALFA \citep{Cordes2006}, Green Bank Northern Celestial Cap \citep{Stovall2014}, and \emph{Parkes} Multibeam Pulsar Survey \citep{Manchester2001}. However, these surveys are inherently biased toward detecting pulsars with short periods (up to tens of seconds), limiting sensitivity to longer-period objects \citep{Lazarus2015,Morello2020,Beniamini2023}. This has contributed to the prevailing paradigm that most radio-emitting NSs must be fast rotators. 
This paradigm was exacerbated by theoretical predictions of a pulsar `death valley'. This is a region in the so-called period $P$ vs. period derivative $\dot{P}$ diagram (see Figure\,\ref{fig:ppdot})
where the pulsar magnetic fields and rotational energy become too weak to sustain the processes required for radio emission, marking the end of their observable activity as pulsars. Most known pulsars lie well outside this region, with short spin periods and moderate spin-down rates.

Recent breakthroughs have challenged this long-standing view. A rapidly growing number of ultra-long period (ULP) radio sources, defined here as likely compact objects exhibiting periodic radio emission with periods $\gtrsim$50\,s, have been detected in the past few years. 
These discoveries prompt a re-evaluation of the physical mechanisms behind coherent radio emission in compact objects and raises intriguing possibilities about their nature. Their properties indicate that some may be highly magnetized white dwarfs (WDs), while others may represent a new population of NSs lying at or near the death valley in the $P$--$\dot{P}$ diagram of pulsars. The existence of ULP sources challenges our understanding of this end-stage in pulsar evolution and suggests that these objects may be more numerous than previously thought.

The Seventeenth Marcel Grossman Meeting provided a platform for enriching discussions on this growing class of sources. This proceeding summarizes the observational properties of the currently known sample while also exploring the models proposed to explain their diverse properties within the compact object population.

\vspace{-0.32cm}
\section{The sample} 
An observational overview of ULP sources was provided by \emph{Kaustubh Rajwade} and \emph{Zorawar Wadiasingh}.
As of November 2024, this emerging class includes 12 members. Three have been identified as WDs, while the nature of several others remains elusive. In the following, we categorize the sources into three classes: {\it i)} WD pulsars; {\it ii)} ULP radio sources of uncertain origin; {\it iii)} the central compact object within the supernova remnant (SNR) RCW~103, which is unique in having no detected radio emission but exhibits magnetar-like behavior (i.e, X-ray bursts and outbursts). 
Table\,\ref{tab:list} lists their observational properties, including the distances estimated using the Galactic free electron density models YMW16 \citep{Yao2017} and NE2001 \citep{Cordes2002}, and the \emph{Gaia} parallax for those sources with a detected optical counterpart.

\begin{table*}[ht]
\begin{center}
\caption{\footnotesize{Properties of ULP sources.}}
\label{tab:list}
\resizebox{\textwidth}{!}{
\begin{tabular}{cccccccc}
\hline
White dwarf 	  & $P_{\rm spin}$   & $\dot{P_{\rm spin}}$ \rule{0pt}{3ex} & $P_{\rm orb}$   & $d$    &  Duty cycle  & $T_{\rm b}$ & $\alpha^a$   \\
pulsars     	  & (s)              & (s\,s$^{-1}$)                        & (hr)            & (pc)   & (\%)         & (K)         &   \\
\hline
AR Sco            & 117.12045378(4) & $6.63\times10^{-13}$$^b$                 & 3.56484672(19)     & 116.7$\pm$0.4$^c$ & $\simeq$50 & 10$^{12}$--10$^{13}$ & --   \\
J1912--4410       & 319.34903(8)    & --                                    & 4.03487736(86)  & 237$\pm$5$^{\dagger}$           & $<1.2$ & --  & $\approx-3$     \\
ILT\,J1101+5521   & 7531.1700(24)   & $\leq3.04\times10^{-11}$              & 2.09199168(7)         & 504$^{+148}_{-109}$ $^{\dagger}$ & $\simeq$2 & 10$^{13}$--10$^{16}$ & -4.1$\pm$1.1   \\
\hline
Ultra-long-$P$     	  & $P_{\rm spin}$  & $\dot{P_{\rm spin}}$ \rule{0pt}{3ex} & DM  		        & $d$    & Duty cycle  & $T_{\rm b}$ & $\alpha^a$  \\
radio sources     & (s)             & (s\,s$^{-1}$)                        & (pc\,cm$^{-3}$)   & (kpc)  & (\%)        & (K) &  \\
\hline
GCRT\,J1745--3009    & 4620.72(1.26)    & --                               & --           & --        & $\simeq$10 & -- & --6.5$\pm$3.4  \\
GLEAM-X\,J1627--5235 & 1091.1690(5) 	& $\leq1.2\times10^{-9}$ 	       & 57$\pm$1 & 1.3$\pm$0.5$^{\ddag}$ & 2--5 & 10$^{16}$--10$^{18}$ & -1.16$\pm$0.04  \\
PSR\,J0901--4046     & 75.88554711(6)   & $(2.25\pm0.1)\times10^{-13}$     & 52$\pm$1 & 0.328$^{\ddag}$, 0.467$^*$ & $\simeq$1 & 10$^{14}$--10$^{15}$ & -1.7$\pm$0.9     \\
GPM\,J1839--10       & 1318.1957(2) 	& $<3.6\times10^{-13}$ 		& 273.5$\pm$2.5 & 5.7$\pm$2.9$^{\ddag}$ & 2.2--22.7 & 10$^{16}$--10$^{20}$ & -3.17$\pm$0.06           \\
ASKAP\,J1935$+$2148  & 3225.309(2) 	    & $\leq(1.2\pm1.5)\times10^{-10}$ & 145.8$\pm$3.5 & 4.3$^{\ddag}$, 5.4$^*$ & 0.3--1.5 & 10$^{14}$--10$^{16}$ & -1.2$\pm$0.1 \\
GLEAM-X\,J0704--37   & 10496.5575(5) 	& $<1.3\times10^{-11}$ 		& 36.541$\pm$0.005 & 0.4$\pm$0.1$^{\ddag}$, 1.8$\pm$0.5$^*$ &  0.3--0.6 & -- & -6.2$\pm$0.6 \\
CHIME\,J0630$+$25   & 421.35542(1) 	& $(-2.5\pm1.6)\times10^{-12}$  & 22$\pm$1 & 0.170$\pm$0.080$^{\ddag}$  &     0.4--0.8 & -- & $\approx-2$ \\ 
J1832--0911         & 2656.247(1) 	& $\leq9.8\times10^{-10}$  & 458--480 & 4.5$\pm$1.2$^{\ddag}$  &     5--10 & 10$^{15}$--10$^{19}$ & $\approx-1.5$ \\ 
\hline
         	& $P_{\rm spin}$      & $\dot{P_{\rm spin}}$      & & $d$    &   &   &   \\
         	& (s)                 & (s\,s$^{-1}$)  & & (kpc)  &   &   &   \\
\hline
1E\,161348--5055$^d$ & 24030.42(2) & $<1.6\times10^{-9}$ & & $\simeq$3.3  &        &    &         \\
\hline
\hline
\end{tabular}
}
\scriptsize{
$^a$ Spectral index of the radio pulsed emission. In many cases, an average value is reported.
$^b$ See \cite{Gaibor2020,Pelisoli2022}.
$^c$ The value is determined using a combination of \emph{VLBI} and \emph{Gaia} astrometry \citep{Jiang2023}.
$^d$ 1E\,161348--5055 is the only source without a radio detection that exhibits magnetar-like behavior. $P$ and $\dot{P}$ were derived during a quiescent state \citep{esposito2011}. The distance refers to that estimated for the associated SNR \citep{caswell1975}.
$^\dagger$ Obtained from the $Gaia$ parallax.
$^{\ddag}$ Obtained from the YMW16 model.
$^*$ Obtained from the NE2001 model.
}
\end{center}
\end{table*}

\vspace{-0.2cm}
\subsection{WD pulsars}

\subsubsection{AR Scorpii}
AR Scorpii (AR Sco) is an M star -- WD binary system in a 3.56-hour orbit, discovered in 2016 \citep{marsh2016}. Unlike standard cataclysmic variable systems, there is no evidence of accretion onto the WD from its companion. Instead, AR Sco exhibits strongly modulated emission across radio to X-rays at a period of 1.97\,min, establishing it as the first example of a WD pulsar \citep{buckley2017}. This periodicity is consistent with the 
beat frequency between the spin period of the WD (1.95\,min) and the binary orbital period, suggesting that the interaction between the compact object and the star drives the pulsed emission. The pulse profile displays two peaks and is aligned across the optical, UV and X-ray energy bands.
The radio emission is weakly linearly polarized and strongly circularly polarized (reaching $\simeq$30\%).  In contrast, the optical emission exhibits strong pulsed linear polarization (up to 40\%), modulated on both the spin and beat periods, and a low-level ($<$a few \%) circular polarization. The double-peaked pulse profile shape and the optical linear polarization morphology are reminiscent of those of the Crab pulsar.
Recent modelling of the optical polarization properties indicates that AR Sco is nearly an orthogonal rotator, with the magnetic inclination angle and observer’s viewing angle varying over the orbital phase \cite{Duplessis2022}. The best fit to the polarization data implies an observer angle of $\approx$70\deg.  

The optical spectra of Ar Sco result from a combination of a blue continuum and the red spectrum of an M-type main-sequence star with atomic emission lines originating near the surface of the star facing the WD. The total radio emission exhibits a power-law spectrum between 1 and 10\,GHz, with a spectral index of $\alpha=0.358\pm0.015$\footnote{The radio flux density scales with the observing frequency as $F_\nu \propto \nu^{\alpha}$.}. The X-ray emission, meanwhile, is dominated by several optically thin thermal plasma emission components with different temperatures, with a power-law signature in the pulsed component. Overall, the broadband spectral and polarimetric properties are consistent with synchrotron radiation from relativistic electrons powered by the spin-down of the magnetized WD.  

\vspace{-0.25cm}
\subsubsection{J191213.72--441045.1}
J191213.72--441045.1 (J1912--4410 henceforth) was discovered during a targeted search for binary WD pulsars \citep{Pelisoli2023}. 
The source was initially flagged due to its variability and its position on the \emph{Gaia} colour-magnitude diagram, and is located at a distance of 237$\pm$5\,pc. Follow-up observations
revealed that J1912--4410 exhibited strong periodic pulses with a 5.3-min period from radio to X-rays (see also \cite{Schwope2023,Pelisoli2024}). 
A narrow pulse is more prominent in the radio band, while a broader component dominates at optical and X-ray wavelengths. 

Optical spectroscopy shows a blue continuum, attributed to emission from the WD, overlaid with the red spectrum of an M-type dwarf star. Strong Balmer and neutral helium emission lines are also observed. Radial velocity analysis reveals a 4-hr orbital period, consistent with a compact binary system. 
The X-ray and UV emissions show a roughly sinusoidal modulation at this period \citep{Schwope2023}.
Modelling of the far UV spectrum gives a WD temperature of 11485$\pm$90\,K and a mass of 0.59$\pm$0.05\,$M_{\odot}$, while the lack of strong Zeeman splitting in the spectral lines places an upper limit on the magnetic field strength of around 50\,MG. The mass of the companion star is determined to be 0.25$\pm$0.05\,$M_\odot$.

X-ray observations revealed variability in the X-ray emission on long (months) and short (hours) timescales. The X-ray spectrum is well described by a combination of a thermal plasma component and a non-thermal power-law component (contributing 90\% of the flux), with a mean X-ray luminosity of $1.4\times10^{30}$\,erg\,s$^{-1}$ \citep{Schwope2023}. 

Similar to AR Sco, the broad-band pulsed emission from J1912--4410 is attributed to synchrotron radiation from the interaction between the WD magnetic field and the M-dwarf companion. Remarkably, while AR Sco shows no signs of ongoing accretion, J1912--4410 exhibits flaring activity, likely due to intermittent accretion, e.g. through wind capture from the companion. This flaring activity supports models suggesting that accretion plays a key role in spinning up the WD and sustaining its magnetic field. The presence of residual accretion suggests that J1912--4410 may be in an earlier evolutionary stage than AR Sco, on its way from an accreting system to a detached binary configuration.

\vspace{-0.25cm}
\subsubsection{ILT\,J1101$+$5521}
ILT\,J1101$+$5521 was discovered in February 2015 through a bright radio pulse detected by the \emph{LOw-Frequency ARray} (\emph{LOFAR}) using short-duration snapshot images \citep{deRuiter2024}. Six additional pulses were identified across five different \emph{LOFAR} observing sessions from 2015 to 2020. These pulses lasted between 30 and 90\,s, displayed significant variability in peak brightness, from 40 to 260\,mJy/beam, and had a duty cycle of 2\%. No further pulses were detected in a 2023 follow-up observation, highlighting the highly variable nature of the source. The pulses were detected across the frequency range of 120–168\,MHz and exhibited a periodicity of approximately 125 minutes, determined via phase-connected timing analysis. The brightest pulse showed a steep spectral index ($\alpha$ = --4.1$\pm$1.1) and strong linear polarization (51$\pm$6\%), suggesting that it was produced in the presence of ordered magnetic fields.

ILT\,J1101$+$5521 is spatially associated with an optical source with a magnitude $r$$\simeq$20.9, located at a distance of approximately 500\,pc. Follow-up spectroscopic observations identified this source as an M4.5V star with significant radial velocity variations, indicative of a binary system. Notably, the inferred binary orbital period matches the 125-min periodicity of the radio pulses. The modelling of broadband photometric data indicated that the M dwarf star (with a mass of $\simeq$0.19\,$M_\odot$) revolves around a likely WD. All the observed properties pointed to a polar-like system where the WD spin period is locked with the binary orbital period. 

Recently, Qu \& Zhang \citep{Qu2024} proposed that the coherent, polarized radio pulses could be produced via a relativistic electron cyclotron maser mechanism occurring along magnetic field lines that connect the WD to the M dwarf star. This process would be triggered by the unipolar induction mechanism, in which the orbital motion of the weakly magnetized M dwarf star through the WD magnetosphere induces an electric potential that powers the emission -- similar to the Jupiter-Io interaction in our solar system. The authors argue that this model could provide a plausible explanation for the coherent radio emissions observed in other ULP radio sources.

\subsection{ULP radio sources}
\subsubsection{GCRT\,J1745--3009: the `Galactic Centre Burper'}
In 2002, the transient radio source GCRT\,J1745--3009 was discovered when 5 powerful bursts at 330\,MHz were detected by the \emph{Karl G. Jansky Very Large Array} (\emph{VLA}) during a monitoring program of the Galactic Centre \citep{Hyman2005}. Originating $\approx$1.25\deg\ south of the Galactic Centre, each burst lasted $\simeq$10 min with about 3\% variation in duration. The bursts reached amplitudes of $\approx$1\,Jy, but no emission was detected between bursts, with upper limits of $\simeq$30\,mJy/beam. The bursts recurred every $\simeq$77 min, following a pattern of steep rise, gradual brightening, and steep decay. They exhibited a very steep average spectral index of $\alpha$ = --6.5$\pm$3.4 \citep{Spreeuw2009}. The brightness temperature, $\simeq$10$^{16}$\,K, suggested a coherent emission mechanism rather than incoherent synchrotron radiation. GCRT\,J1745--3009 was re-detected at 330\,MHz in 2003 and 2004 using the \emph{Giant Metrewave Radio Telescope} (\emph{GMRT}). In 2003, the burst had a maximum flux density of $\simeq$0.5\,Jy and a decay time of $\approx$2 min, consistent with the 2002 bursts \citep{Hyman2006}. Significant detection of circularly polarized emission unequivocally indicated a coherent emission process \citep{Roy2010}. The 2004 burst was much fainter, $\simeq$50\,mJy, lasted only $\simeq$2 min, and had an even steeper spectral index of $\alpha=-13.5\pm3.0$ compared to the 2002 events \citep{Hyman2007}. The source exhibits detectable bursting activity about 7\% of the time \citep{Hyman2007}. The absence of a counterpart in the optical and near-infrared bands \citep{Kaplan2008} suggests that GCRT\,J1745--3009 is likely located far away, possibly $\simeq$4\,kpc \citep{Roy2010} or even near the Galactic Centre at $\approx$8.5\,kpc.

Several scenarios have been proposed to explain the nature of GCRT\,J1745--3009. One model involves a double NS binary system \citep{Turolla2005}, where radio emission is produced by particles accelerated at a shock formed by the interaction between the pulsar wind and the magnetosphere of its companion as it approaches periastron in an eccentric 77-min orbit. Alternatively, it might be an unusual magnetar with a much longer period than typical \citep{Hyman2005}, though the very steep spectral index and circular polarization observed make this scenario less likely. Another possibility is electron cyclotron maser emission or plasma emission from a magnetized subsolar dwarf with a 77-min spin period at a distance of $\simeq$4\,kpc \citep{Roy2010}. Finally, Zhang \& Gil \citep{Zhang2005} suggested it could be a transient, strongly magnetized WD pulsar with the 77-min cycle being its rotation period, and the 10-min flares caused by the radio beam sweeping past our line of sight. The bursts would occur when stronger magnetic fields emerge in the WD polar cap region, triggering radio pulsar-like emission; when these conditions end, the flares stop, and the WD enters a `dormant' phase.

\vspace{-0.25cm}
\subsubsection{GLEAM-X\,J162759.5--523504.3}
GLEAM-X\,J162759.5--523504.3 (GLEAM-X\,J1627--5235) was identified during an image-plane survey of the Galactic plane conducted using the \emph{Murchison Widefield Array} (\emph{MWA}) \citep{Hurley-Walker2022}. This source exhibits periodic radio pulses at a period of 1091\,s -- much longer than the rotation periods of known pulsars and magnetars.

The discovery took place during two active intervals between January and March 2018, during which 71 distinct pulses were observed. These pulses had flux densities ranging from 5 to 40\,Jy and lasted between 30 and 60\,s. The pulse structures varied significantly, sometimes appearing as complex, spiky bursts or subdivided into smaller sub-pulses. Spectral analysis revealed a dispersion measure (DM) of 57$\pm$1\,pc\,cm$^{-3}$, corresponding to a distance of 1.3$\pm$0.5\,kpc according to the YMW16 model \cite{Yao2017}. The radio spectrum showed a spectral index of $\alpha$= --1.16$\pm$0.04, indicative of non-thermal processes. Additionally, the pulses exhibited a remarkably high degree of linear polarization at (88$\pm$1)\%, with no significant variation in polarization angle over the observation period. The Faraday rotation measure was consistent with expectations for the Galactic region where the source is located. Deep X-ray observations did not detect any X-ray emission from the source during a radio-inactive period, with an upper limit on the X-ray luminosity of 10$^{29}$--10$^{30}$\,erg\,s$^{-1}$ depending on the assumed spectral model \citep{Rea2022}. Additionally, optical and near-infrared data from various public surveys found no evidence of an optical counterpart within the positional uncertainty of the radio source \citep{Rea2022}.

The emission properties of GLEAM-X\,J1627--5235 are inconsistent with known phenomena such as flare stars, exoplanets, or typical pulsars; instead, the regularity and high linear polarization of the pulses suggest the presence of a strongly magnetized object, favouring an interpretation as a magnetar. However, the absence of detectable X-ray emission, typically associated with magnetars, leaves alternative interpretations open (for more extensive discussions, see Sections\,\ref{sec:gleamx_scenarios} and \ref{sec:fallback}).

\vspace{-0.25cm}
\subsubsection{PSR\,J0901--4046}
The radio-emitting source PSR\,J0901--4046 was serendipitously discovered by \emph{MeerKAT} in September 2020 \citep{Caleb2022}.
In this first observation, 14 pulses were identified regularly spaced over a time span of $\simeq$30 min, repeating with a period of $P\simeq$76\,s. Follow-up observations were carried out between September 2020 and May 2021. Single pulses were detected at every rotation of the source with a very narrow duty cycle ($\simeq1\%$) and stable pulse profiles. The pulses showed unique spectro-temporal properties, such as quasi-periodicity and partial nulling.
 
The monitoring campaign, spanning $\simeq$7 months, allowed for an estimate of the period derivative ($\dot{P}\simeq2.25\times10^{-13}$\,s\,s$^{-1}$) and thus a characteristic age of $\simeq5.3$\,Myr, a surface dipolar magnetic field strength of $\simeq1.3\times10^{14}$\,G and a spin-down luminosity of $\simeq2\times10^{28}$\,erg\,s$^{-1}$ (assuming that the NS undergoes dipolar losses exclusively). The measured DM of 52$\pm$1\,pc\,cm$^{-3}$ corresponds to a distance of $\simeq$0.3\,kpc and $\simeq$0.5\,kpc according to the YMW16 \citep{Yao2017} and NE2001 \citep{Cordes2002} models, respectively. The deepest image of the field revealed diffuse emission around PSR\,J0901--4046, though a deeper analysis is required to claim the association between the two. If the diffuse emission were ascribed to the SNR from the event that formed the NS, it would suggest that the source is younger than the characteristic age.  

While PSR\,J0901--4046 could be an old magnetar, no X-ray emission was detected and the L-band in-band spectral index ($\alpha$= --1.7$\pm$0.9) is closer to the values found for rotation-powered pulsars than those measured for radio-loud magnetars ($\alpha \simeq -0.5$). Moreover, the source lies far from the known magnetar population in the pulsar $P-\dot{P}$ diagram. On the one hand, it stands below the `death line' of the vacuum-gap curvature radiation models \cite{Zhang2000}, according to which pulsars in this region of the $P-\dot{P}$ diagram cannot support pair-cascade production above the polar cap in their inner magnetosphere, leading to the cessation of radio emission. 
On the other hand, it lies above the `death line' of the space-charge-limited flow radio-emission model \cite{Zhang2000}, where pair cascades are driven by non-relativistic charges flowing freely from the polar cap, assuming a multipolar magnetic field configuration. Alternatively, PSR\,J0901--4046 could be a radio pulsating WD binary system, though multi-wavelength observations did not find evidence to support this hypothesis.


\vspace{-0.25cm}
\subsubsection{GPM\,J1839--10}
GPM\,J1839$–$10 was first detected during a monitoring campaign of the Galactic plane using the \emph{MWA} \citep{Hurley-Walker2023}. The initial discovery occurred when two significant 30-s-wide pulses were detected in a single night, prompting further monitoring with several radio telescopes, including the \emph{Australia Telescope Compact Array}, the \emph{Parkes}/\emph{Murriyang} radio telescope, the \emph{Australian Square Kilometre Array Pathfinder} (\emph{ASKAP}), and \emph{MeerKAT}. In particular, \emph{MeerKAT} interferometric observations precisely pinpointed the position of GPM\,J1839$–$10 and allowed for a detailed analysis of its pulse structure. The derived DM was found to be 273.5$\pm$2.5\,pc\,cm$^{-3}$, converting to a distance of 5.7$\pm$2.9\,kpc according to the YMW16 model \citep{Yao2017}. The detected pulses varied in width from 0.2 to 4\,s and exhibited remarkable polarization features, including orthogonal polarization mode transitions typically seen in NS pulsars. These pulses also displayed quasi-periodic oscillations, similar to pulsar microstructure albeit on a much larger scale.

Archival searches revealed that GPM\,J1839$–$10 had been active since at least 1988, with detections at various radio telescopes such as the \emph{VLA} and the \emph{GMRT}. The observed light curves showed substantial variability in pulse flux (with flux densities ranging from 0.1 to 10\,Jy), widths (between 30 and 300\,s) and shape. Remarkably, the source exhibited a significant nulling fraction, with 50–70\% of pulses disappearing below detectability. No persistent radio emission was detected outside pulse windows, down to a brightness upper limit of 60 mJy/beam (3$\sigma$).

Analysis of the pulse arrival times over nearly 34 years revealed a rotational period of 1318\,s. The spin period derivative was constrained to $\dot{P}<3.6\times10^{-13}$\,s\,s$^{-1}$, suggesting the source lies just below the death valley on the $P$--$\dot{P}$ diagram.

The radio luminosity of GPM\,J1839$–$10 is unusually high, even surpassing the typical spin-down luminosity in NS scenarios. This behavior is reminiscent of the single radio pulses seen in magnetars, challenging conventional models of pulsar emission. The long-lived and intermittent activity of GPM\,J1839$–$10 over at least three decades is also highly unusual. Unlike GLEAM-X\,J1627--5235, which exhibited only brief radio activity, GPM\,J1839$–$10 remains active despite its long period and high nulling fraction. Additionally, simultaneous X-ray observations failed to detect any counterpart, ruling out a direct connection to magnetar-like X-ray activity. Alternative explanations, such as a highly magnetic isolated WD, star-exoplanet interactions or brown dwarf binaries, have also been considered though they do not fully account for all the unique characteristics of GPM\,J1839$–$10.

\vspace{-0.35cm}
\subsubsection{ASKAP\,J193505.1$+$214841.0}
During a target of opportunity observation of a gamma-ray burst, \emph{ASKAP} found a new source, ASKAP\,J193505.1$+$214841.0 (ASKAP\,J1935$+$2148), through the detection of four bright and tens-of-seconds-wide pulses with a $>$90\% degree of linear polarization \citep{Caleb2024}. Follow-up observations were carried out with \emph{MeerKAT}, whose better timing resolution allowed for an estimate of the DM (145.8$\pm$3.5\,pc\,cm$^{-3}$). The single pulses detected with \emph{MeerKAT} displayed different properties from those observed with \emph{ASKAP}: a width of $\simeq$370\,ms and circular polarization fraction exceeding 70\%. The period of $\approx$3225\,s ($\simeq$54 min) and the periode derivative upper limit ($\dot{P}\leq(1.2\pm1.5)\times10^{-10}$\,s\,s$^{-1}$) place ASKAP\,J1935$+$2148 in the pulsar `death valley' in the $P-\dot{P}$ diagram, like other ULP radio sources, challenging current models of radio emission. 
ASKAP\,J1935$+$2148 showed three different emission states: 
{\it (i)} one characterized by highly linearly polarized bright pulses with widths of 10--50\,s, as observed with ASKAP;  
{\it (ii)} one exhibiting highly circularly polarized weak pulses, which are about 30 times fainter than the bright ones and have a milliseconds-width, as seen with \emph{MeerKAT};
{\it (iii)} a quiescent state with no detectable pulses, observed with both telescopes.
ASKAP\,J1935$+$2148 is the only known ULP source to show three distinctive emission modes, resembling mode-switching pulsars.

Several X-ray observations covered the field of ASKAP\,J1935$+$2148. The X-ray luminosity upper limit, $L_X<4\times10^{30}$\,erg\,s$^{-1}$, is consistent with the values derived for other ULP radio sources. Moreover, archival near-infrared images showed a catalogued source within the positional circle of ASKAP\,J1935$+$2148. Spectral properties and magnitudes suggest that this object is an L/T-dwarf star, and the \emph{Gaia} parallax implies a distance of $<$0.5\,kpc for such a star, which would be otherwise undetectable at 4.85\,kpc. Hence, it is most likely a foreground star, unassociated with ASKAP\,J1935$+$2148. A constraint on the source radius for the observed period rules out an isolated WD; a cataclysmic variable scenario cannot explain the high radio luminosity of the source. Hence, ASKAP\,J1935$+$2148 is much more likely to be an ULP magnetar or a very peculiar NS, either isolated or in a binary system.  

\vspace{-0.25cm}
\subsubsection{GLEAM-X\,J0704--37}
The discovery of GLEAM-X\,J0704$-$37 was made using the GaLactic and Extragalactic All-sky \emph{MWA} (GLEAM-X) survey \citep{Hurley-Walker2024}. The source was first detected as a 30-s-wide pulse during a 2-min observation. Subsequent archival searches revealed 33 similar pulses, with the shortest interval between consecutive pulses being approximately 2.9 hours. This led to the identification of the source periodicity. Further radio follow-up with the \emph{MeerKAT} telescope confirmed the periodic nature of GLEAM-X\,J0704$-$37, detecting pulses at a period consistent with the initial discovery. Spectral analysis of the radio emission suggests a highly structured emission with significant levels of linear polarization (20--50\%), indicating an environment with strong and ordered magnetic fields. Timing residuals from pulse arrival times indicated an additional ULP sinusoidal modulation at a period of roughly 6.23 years. This could suggest either a long-term orbital period or a more complex internal process, such as timing noise.
An optical counterpart was identified using archival data. The counterpart was classified as a cool main sequence star of spectral type M3V based on spectral analysis. The estimated distance to the system, based on dispersion measures and \emph{Gaia} parallax data, is about 1.5\,kpc, though this value remains uncertain. Deep X-ray observations did not detect any emission from GLEAM-X\,J0704$-$37, setting an upper limit on its X-ray luminosity of a few $10^{30}$\,erg\,s$^{-1}$. This non-detection, combined with the radio data, supports the idea that the source might be a faint, old NS or a WD. 

The presence of the M3V dwarf star suggests that GLEAM-X\,J0704$-$37 could be part of a binary system. Scenarios where the observed radio emission could be produced by a slowly rotating NS or a highly magnetized WD have been considered. However, both cases present significant challenges to current models, particularly given the system ULP nature and the complex radio emission patterns observed. If the compact object were a NS, it would be an unusually slow rotator with a weak magnetic field, making it difficult to reconcile with traditional pulsar models. On the other hand, if the companion were a WD, it could be similar to other known systems where the WD produces pulsar-like emissions, although the observed 6.23-years periodicity would still require unconventional explanations.

\vspace{-0.25cm}
\subsubsection{CHIME\,J0630$+$25}
CHIME\,J0630$+$25 was discovered using the \emph{Canadian Hydrogen Intensity Mapping Experiment} (\emph{CHIME}) telescope \citep{Dong2024}. Seventeen bursts were detected over an extended period of observations. These bursts had fluences in the range of 60--670\,Jy\,ms at 400\,MHz, widths of up to 4\,s and an average spectral index of $\alpha\simeq-2$. The pulses exhibited complex temporal and spectral structures, with some displaying substructures reminiscent of those seen in radio-loud magnetars. A phase-coherent timing solution revealed a period of around 421\,s, making it one of the longest period radio emitters identified to date. However, the derived period derivative $\dot{P}=(-2.5\pm1.6)\times10^{-12}$\,s\,s$^{-1}$ suggests that the source may be spinning up, which is unusual for isolated NSs and raises the possibility that CHIME\,J0630$+$25 could be part of a binary system. X-ray observations identified four potential X-ray counterparts within the localization region of CHIME\,J0630$+$25, but none of these sources were conclusively linked to the radio source. The source distance was estimated at 170$\pm$80\,pc, assuming the YMW16 model for the Galactic electron density \cite{Yao2017}.

Both NS and WD models have been considered to explain the nature of CHIME\,J0630$+$25. The sporadic and complex nature of its radio emission draws parallels with radio-loud magnetars. However, the observed period is much longer than those of NSs, leading to considerations of a WD origin. WDs with similar spin periods have been observed to emit pulsed radio signals, though the luminosity of CHIME\,J0630$+$25 is notably higher than that of known WD pulsars.

The proximity of CHIME\,J0630$+$25 makes it a prime candidate for further multiband follow-up studies, which could provide additional insights into its nature and determine whether it aligns more closely with known NSs or WDs.

\subsubsection{J1832--0911}
J1832--0911 is a newly discovered ULP source spatially associated with the SNR G22.7$-$0.2, independently discovered by the DAocheng Radio Telescope (DART) and the ASKAP, exhibiting a period of $\simeq$44 min \cite{Wang2024,Li2024}. Initial DART observations detected periodic pulses with peak flux densities between 0.5 and 2\,Jy across frequencies of 149--459\,MHz. A subsequent FAST observation at L-band captured a short, nearly 100\% linearly polarized pulse lasting 0.2\,s, suggesting a coherent emission mechanism similar to that of pulsars \cite{Li2024}. 
ASKAP observations revealed a pulse lasting $\simeq$2 min with a peak flux density of 1.87\,Jy, spectral index $\alpha = -1.5 \pm 0.1$ and a total fractional polarization of $92 \pm 3\%$. Further observations with various radio telescopes detected multiple periodic pulses from J1832--0911 with diverse properties in terms of morphology, flux density ($\approx$30\,mJy -- 20\,Jy), spectral index ($-2.2$ and $-0.3$), and polarization \cite{Wang2024}. A DM of $\approx$458--480\,pc\,cm$^{-3}$ was measured, corresponding to a distance of $\simeq$4.5\,kpc, consistent with that of the SNR G22.7$-$0.2.

Despite precise localization, no optical or infrared counterparts were detected. Archival X-ray data from {\it XMM-Newton} did not reveal any emission, but a serendipitous detection in February 2024 by {\it Chandra} showed an X-ray counterpart with a period consistent with the radio period and a luminosity of $\approx7.4 \times 10^{32}$\,erg\,s$^{-1}$. This X-ray emission decreased significantly about half a year later, mirroring a decline in radio flux, indicating correlated variability across wavelengths. Similar to other ULP sources, J1832--0911 is located in the death valley of the $P$--$\dot{P}$ diagram. 

J1832--0911 was proposed to be a NS that has been rapidly spun down through interaction with a fallback disc from the SN explosion. This mechanism could account for the ultra-long rotation period within $10^4-10^5$ years after the SN explosion, aligning with the estimated age of SNR G22.7$-$0.2. Alternatively, it could be an unusual magnetar with a core-dominated magnetic field configuration. However, the low spin-down rate and quiescent X-ray luminosity present challenges to existing magnetar models, which typically require higher magnetic fields and associated X-ray emissions. A third possibility is that J1832--0911 is a WD binary system. However, the required magnetic field strength (exceeding $5 \times 10^9$\,G) and the lack of optical counterparts make this scenario less plausible. Additionally, the energy output observed exceeds what is typically seen in WD systems.

\subsection{A magnetar-like object with a 6.67-hr period}
\label{sec:rcw} 
During the session, \emph{Alice Borghese} presented the discovery and findings on 1E\,161348--5055 (1E\,1613). Discovered close to the centre of the young ($\simeq$2\,kyr) SNR RCW\,103, 1E\,1613 was identified as the first radio-quiet, X-ray-bright isolated NS in a SNR\cite{tuohy1980}. 
Historically labelled as a central compact object (CCO), 1E\,1613 showed properties that built a unique phenomenology in the NS scenario. Firstly, at odds with the other CCOs, it is strongly variable, displaying flux variations on time scales of months/years. In 1999, a large outburst occurred with the flux increasing by a factor of $\simeq$100. Secondly, a long \emph{XMM-Newton} observation, which caught the source in a low state, yielded the unambiguous evidence of a periodic modulation at 6.67$\pm$0.03\,hr\citep{deluca2006}. The corresponding pulse profile changes according to the flux level from sine-like shape when the source is in a low state (observed 0.5--8\,keV flux of $\simeq10^{-12}$\,erg\,s$^{-1}$\,cm$^{-2}$) to more complex, multi-peaked configurations in high state ($\simeq10^{-11}$\,erg\,s$^{-1}$\,cm$^{-2}$). Based on these characteristics, 1E\,1613 was proposed to be either the first low-mass X-ray binary in a SNR with an orbital period of 6.67\,hr or a young isolated magnetar with a 6.67-hr spin period. 

In 2016, new events shed light on the nature of the source: a short magnetar-like burst of hard X-rays was detected from the direction of RCW\,103 and, simultaneously, a flux enhancement of 1E\,1613 by a factor $>$100 with respect to the quiescent level, attained up to one month before, was observed \cite{rea2016}. A few days after the outburst onset, {\it Chandra} and {\it NuSTAR} observations were performed. For the first time, the source was detected at hard X-rays up to $\simeq$30\,keV; this emission was related to a non-thermal spectral component described by a power law. The light curve exhibited two peaks per phase cycle, which is diﬀerent from the sinusoidal shape seen during the quiescent state. Moreover, the outburst prompted searches for an infrared counterpart. The \emph{Hubble Space Telescope} and \emph{Very Large Telescope} images disclosed a new object, not detected in archival observations\cite{tendulkar2017,esposito2019}. The counterpart properties ruled out the binary scenario, favouring the isolated object interpretation. As a matter of fact, the X-ray-to-infrared luminosity ratio of $\simeq10^5$ is consistent with typical values or limits for magnetars and isolated NSs. 1E\,1613 was intensively monitored in the following years and reached quiescence in late 2018.

All the aspects caught by these observations (i.e., the discovery of a non-thermal hard spectral component at the outburst peak, the pulse profile variability in time, the flux decay pattern) can be naturally explained if 1E\,1613 is a magnetar. However, its long periodicity makes this object unique among the known magnetars, whose rotational periods are smaller by three orders of magnitude. A very eﬃcient braking mechanism is required to slow down the source in $\simeq$2\,kyr from its birth period to the currently value. Most models consider a propeller interaction with a fallback disc that can provide an additional spin-down torque. For example, Ho \& Andersson\cite{ho2017} predicted a remnant disc of 10$^{-9}$\,M$_{\odot}$ around a rapidly rotating NS that is initially in an ejector phase; after hundreds of years, the NS rotation becomes slow enough to allow the onset of a propeller phase. To reproduce the observational properties, 1E\,1613 should have a slightly higher surface dipolar component of the magnetic field than known magnetars, $\simeq 5\times10^{15}$\,G.

\vspace{-0.25cm}
\section{Nature of ULP sources: lessons from GLEAM-X\,J1627--5235}
\label{sec:gleamx_scenarios}
Scenarios for the nature of these sources have been discussed by \emph{Zorawar Wadiasingh}. Here, we focus on the two possibilities proposed for GLEAM-X\,J1627--5235.

\vspace{-0.25cm}
\subsection{Magnetar scenario}
In the magnetar hypothesis, the 18-min spin period suggests GLEAM-X\,J1627--5235 underwent an unusual evolutionary process that slowed its rotation over time. A likely mechanism for this slowdown could be supernova fallback accretion. In this scenario, material from the supernova that formed the NS falls back onto it, adding mass and angular momentum, which exerts a braking torque that slows the star spin (see Sect.\,\ref{sec:fallback}). Additionally, episodes of enhanced spin-down, possibly driven by mass-loaded particle winds and outflows following giant flares, could contribute to this effect \citep{Beniamini2023} (for additional braking mechanisms, see \cite{Zhou2024}). 

It has been suggested that GLEAM-X\,J1627--5235 is a very old NS, where a strong magnetic field is maintained over millions of years and thus decays very slowly. 
Magneto-thermal simulations suggest that GLEAM-X\,J1627--5235 is $\gtrsim$1\,Myr old for any reasonable crustal magnetic field ($B > 10^{13}$\,G), unless its magnetic field is anchored in the star core, or the star has experienced rapid cooling \citep{Rea2022,Beniamini2023}. Although the transient, coherent, and highly polarized emission are characteristics similar to those observed in radio-loud magnetar, the long spin period and the lack of detectable X-ray emission challenge challenge current magnetar models, which usually predict shorter periods and brighter thermal X-ray emission \citep{Pons2013,Vigano2013}. 

Suvorov \& Melatos \cite{Suvorov2023} provide a detailed evaluation of the magnetar hypothesis for GLEAM-X\,J1627--5235. They conducted simulations to examine the decay of the crustal magnetic field including processes such as Hall–Ohm diffusion and plastic flows. They found that a polar magnetic field strength of at least $5\times10^{15}$\,G is likely necessary for the star to operate as a radio pulsar, based on typical assumptions about the pulsar `death valley'. By tracing the magnetic field decay backward, they argue that the star likely had significant angular momentum at birth, which helped amplify its magnetic field. 
The authors also note that if the object were indeed a young, isolated magnetar -- between 10 and 50\,kyr old -- with a current magnetic field of approximately $10^{16}$\,G, the upper limit on its thermal luminosity would suggest that it is cooling primarily through a direct Urca mechanism.

\vspace{-0.25cm}
\subsection{WD scenario}
If GLEAM-X\,J1627--5235 were a NS, its radiated energy would exceed its spin-down power by orders of magnitude. However, this inconsistency is resolved if the object is instead a WD pulsar, since WDs possess much larger moments of inertia, allowing them to store and radiate greater amounts of energy. Although WDs can also have strong magnetic fields and produce pulsed emissions, the coherent pulses observed in GLEAM-X\,J1627--5235 differ from the incoherent radio emission from AR Sco. This difference could indicate that GLEAM-X\,J1627--5235 is an isolated magnetic WD, rather than part of a binary system like AR Sco. In this framework, its emission may be driven by its own magnetic field rather than by interactions with a companion star. 
On the one hand, a highly magnetized ($\simeq$10$^8$\,G) hot proto-WD pulsar with a mass of $\simeq$0.5\,$M_{\odot}$, a radius of $\simeq$0.3\,$R_{\odot}$, and an age of $\simeq$30\,kyr could, in principle, explain the observed properties of GLEAM-X\,J1627--5235 \citep{Loeb2022}. However, this scenario is disfavoured due to the lack of an optical counterpart. On the other hand, a cold, isolated magnetic WD would be consistent with the X-ray luminosity upper limits \citep{Rea2022}. The Lorentz factor of the radiating particles is estimated to be at least 300, based on the width of the emitted pulses. If the radiation is in the form of curvature radiation, where charged particles are accelerated along curved magnetic field lines, then the inferred radius of curvature of the magnetic field lines would be more consistent with the size of a WD’s magnetosphere than that of a NS’s \citep{Katz2022}.

\vspace{-0.25cm}
\section{Evolutionary pathways: the role of fallback discs}
\label{sec:fallback}
Recently, several studies have started with the assumption that ULP sources are NSs and revisited the role of fallback accretion from the supernova debris in shaping their properties\citep{Gencali2022,Ronchi2022,Gencali2023,Fan2024}.
An overview of these investigations was provided by \emph{Zorawar Wadiasingh}, with a detailed fallback disc model presented by \emph{Ali Arda Gen\c{c}ali}.

The process of fallback accretion can be divided into distinct phases: shortly after the supernova, a powerful neutrino-driven wind from the NS may clear away much of the fallback material. However, some material can collapse back onto the NS, forming a fallback disc. This disc interacts with the stellar magnetic field, potentially imparting a strong torque that significantly slows the star rotation. If the NS magnetosphere extends beyond the corotation radius, the inflowing material can be propelled away -- a process that is especially effective at spinning down the NS. NSs with fallback discs may cause the magnetic field lines to open up, enhancing the number of open field lines at the poles \cite{Parfrey2016,Parfrey2017} and possibly triggering radio pulsations \cite{Gencali2021,Gencali2022,Gencali2023}. In some cases, this fallback accretion leads to a state of spin equilibrium, where the accretion torque balances dipolar radiation losses.  

Recent studies suggest that the combination of fallback accretion and magnetic field decay can lead to the formation of pulsars with spin periods much longer than expected from standard dipolar losses. Models consider various initial conditions, such as the magnetic dipole field strength, birth spin period, and initial fallback accretion rates. Ronchi et al. \cite{Ronchi2022} found that spin periods $\gtrsim$100\,s can be achieved with initial magnetic fields $\gtrsim$10$^{14}$\,G and moderate initial fallback accretion rates ($\simeq10^{22}-10^{27}$\,g\,s$^{-1}$) at relatively young ages (10$^3$ -- 10$^5$ years). Fan et al. \cite{Fan2024} found similar results, but also noted that it is unlikely that GPM\,J1839$-$10 could reach the observed period derivative upper limit during the active lifetime of the fallback disc. Instead, they proposed that this source may have transitioned into a second ejector phase after the fallback disc became inactive, yielding a very slow spin evolution. 

Prior to the discovery of ULP sources, Benli \& Ertan \cite{Benli2016} had predicted their existence through simulations of the long-term evolution of magnetars with relatively strong dipole magnetic fields. In their fallback disc model, ULP sources and other isolated NS populations could be accounted for by invoking conventional dipole fields ($\simeq$10$^{12}$\,G) of young NSs. Using this model, Gen\c{c}ali et al. \cite{Gencali2022} found two possible evolutionary pathways for GLEAM-X\,J1627$–$5235. In one scenario, the fallback disc remains active with ongoing low-rate accretion, continuing to slow the pulsar spin period to a few thousand seconds at a rate of $\dot{P}\simeq10^{-10}$\,s\,s$^{-1}$ until the disc becomes inactive. Alternatively, the disc may have already become inactive, leaving the NS to spin down through magnetic dipole torque alone, resulting in a small spin period derivative ($\dot{P}\simeq4\times10^{-18}$\,s\,s$^{-1}$). In a following paper and again using the same above model, Gen\c{c}ali et al. \cite{Gencali2023} estimated an age of $\simeq(6-8)\times10^{5}$ years for PSR\,J0901$-$4046. More recently, Gen\c{c}ali \& Ertan \cite{Gencali2024_link} showed that the rotational properties of ULP sources, along with those of other isolated NS populations, can be accounted for within a unified framework using more realistic model calculations \cite{Ertan2021}. In this model, most ULP sources are currently evolving in the propeller phase, where the source X-ray luminosity is attributed to cooling luminosity, consistent with the observed luminosity upper limits for these sources. 

More recently, Yang et al. \citep{Yang2024} used Monte-Carlo simulations to investigate the evolution of magnetars' magnetic fields, spin periods, and inclination angles in the presence of fallback discs. They found that thermal viscous instabilities within the disc play a critical role in ULP formation by determining the disc's lifetime. The simulations indicate that many ULPs are likely to be nearly aligned or orthogonal rotators. These insights emphasize the crucial interplay between fallback disc dynamics and magnetic configurations in shaping the properties of ULPs.

\vspace{-0.25cm}
\section{A mechanism for the radio emission}
\emph{Alex Cooper} delivered a talk on mechanisms driving radio emission in ULP sources. Cooper \& Wadiasingh \citep{Cooper2024} have recently suggested that the radio pulses are produced through a coherent emission process within the magnetosphere of a strongly magnetized NS. These sources could be much older than known magnetars, with B-fields persisting for thousands to tens of thousands of years -- far longer than standard magnetar evolution models predict. This suggests ULP sources represent a later stage in NS evolution, where strong B-fields remain active despite slower rotation.

A key aspect of their work is the role of mild, long-lived twists in localized magnetospheric regions powering emission. These twists likely result from slow, plastic deformations in the NS crust, or ``plastic flow'', occurring when magnetic stresses exceed the crust's elastic limits. Additionally, twisting could result from thermoelectric effects caused by temperature gradients within the crust, potentially affecting the long-term evolution of magnetic fields and emissions in older NSs. The dissipation of these magnetic twists may drive the observed electromagnetic radiation, with the rate depending on factors like local magnetic field strength, crustal movement velocity, and the size of the twisted region.

Two mechanisms are proposed for producing the observed radio emission: curvature radiation and resonant inverse Compton scattering (RICS). Both involve particle acceleration in gaps within the twisted magnetosphere, generating cascades of electron-positron pairs that produce coherent radio waves. In curvature radiation, highly relativistic particles move along curved magnetic field lines, emitting photons. In RICS, lower-energy thermal photons scatter off relativistic electrons, resulting in high-energy photons that generate electron-positron pairs. Both processes require magnetar-like magnetic fields ($\gtrsim10^{14}$\,G) and long spin periods, associated with highly ordered magnetic fields likely explaining the polarized radio signals observed.
In this framework, radio pulses are often expected to be double-peaked due to the cone-shaped emission pattern from twisted magnetic fields. Pulse profiles are predicted to be highly variable, depending on crustal plastic motion dynamics and magnetic field configuration, with variability on timescales from sub-ms fluctuations to long-term changes related to crustal motion and magnetic field evolution.

ULP radio sources are expected to produce thermal X-ray/UV emissions from heated NS surface regions due to returning currents from the twisted magnetosphere. Deep observations during periods of radio activity could reveal such thermal emission associated with twisted B-fields and dissipation processes in the NS crust.

\vspace{-0.55cm}
\section{Unveiling the population of ULP radio sources} 
The implications of population synthesis approaches and local density estimations for ULP sources were discussed by \emph{Zorawar Wadiasingh}. Following the discovery of GLEAM-X\,J1627--5235 and PSR\,J0901-4046, it was argued that these objects are most likely highly magnetized sources, consistent with ULP magnetars \citep{Beniamini2023}. 
The local density of objects with properties similar to these sources can be estimated by assuming that each represents the closest member of its class. Adopting a population synthesis approach and assuming a NS nature for these newly discovered sources,  their proximity suggests that there may be thousands to tens of thousands of similar objects in the Galactic field. This imposes constraints on their formation rate and typical ages. GLEAM-X\,J1627--5235 and PSR\,J0901--4046 are estimated to be $\sim10^5-10^6$\,yr old, older than the known Galactic magnetar population. ULP magnetars may thus represent a distinct evolutionary channel from `canonical' magnetars, having undergone a unique evolution of their poloidal magnetic fields.

Recently, Rea et al. \cite{Rea2024} employed population-synthesis simulations to examine the birth rates and evolution of ULP sources, taking into account factors like initial spin-period distributions, masses, radii, beaming angles, and magnetic field configurations. These simulations indicate that it is highly unlikely for long-period NSs to retain enough spin-down energy ($\dot{E}>10^{27}$erg\,s$^{-1}$) to power radio emission at the luminosity levels observed in GPM\,J1839--10. In contrast, long-period WDs might be relatively more common. Nevertheless, no current model can readily account for the observed bright coherent radio emissions in either scenario.

\vspace{-0.25cm}
\section{Concluding remarks and looking forward}
Throughout the Seventeenth Marcel Grossman Meeting, presentations highlighted how ULP (radio) sources are reshaping our understanding of compact objects, their evolution, and emission mechanisms. Despite decades of radio surveys, it became evident that much of the parameter space -- especially the faint and ULP end -- remains largely uncharted.
Participants discussed how current \citep{Shimwell2017,Bhat2023} and upcoming \citep{Keane2015} surveys, such as those using longer dwell times and image-domain approaches \citep{Fender2016,deRuiter2024}, alongside ongoing upgrades to existing facilities, are poised to 
reveal a far larger population of ULP sources. Moreover, improved detection algorithms \citep{Morello2020} applied to archival data may also uncover previously undetected sources \citep{Singh2022,Grover2024}.

A recurring theme during the session was the need for coordinated observations across the electromagnetic spectrum to unravel the true nature of ULP radio sources. For example, while radio observations have provided valuable insights into their coherent emission processes, the absence of detectable X-rays in many instances raises questions about the underlying magnetospheric structure and thermal properties. Multi-wavelength data will be key to probing these aspects, potentially uncovering new properties of their emission that radio studies alone cannot reveal.

From an evolutionary perspective,  the session shed light on alternative and unusual pathways for compact objects. NSs with periods far beyond the conventional values challenge our current understanding of spin-down mechanisms. Theories involving magnetic field decay and interactions with fallback discs suggest these mechanisms may play a crucial role in slowing these stars over timescales much longer than previously anticipated. While these mechanisms are not yet fully understood, they could hold the key to explaining how magnetars and pulsars transition into ULP radio emitters. Advanced simulations of fallback accretion, magnetic-field evolution, and population synthesis studies will help clarify the origins of these sources.
As the population of ULP radio sources grows, existing emission models will need to adapt to account for their slower rotation and unique magnetospheric conditions. 

With new discoveries on the horizon, the coming years hold great promise for deepening our understanding of the physics governing these fascinating sources.

\vspace{-0.25cm}
\section*{Acknowledgements}
This proceeding serves as a summary of the topics discussed at the `Slowly rotating pulsars' session of the Seventeenth Marcel Grossman meeting. 
We thank all participants for their valuable contributions and for engaging in fruitful discussions. We thank Ali~Arda~Gen\c{c}ali, Zorawar~Wadiasingh and Alex Cooper for comments on the manuscript, and Natasha~Hurley-Walker, Nanda~Rea, Kaustubh Rajwade, Emilie~Parent, Vanessa~Graber, Michele~Ronchi and Celsa~Pardo~Araujo for valuable collaborations and discussions on the topic over the past years. FCZ is supported by a Ram\'on y Cajal fellowship (grant agreement RYC2021-030888-I). AB is supported by a L'Oreal--Unesco For Women In Science Fellowship (2023 Italian program).

\vspace{-0.4cm}
\bibliographystyle{ws-procs961x669}

\end{document}